# Origin of robust interaction of spin waves with a single skyrmion in perpendicularly magnetized nanostripes

Junhoe Kim and Sang-Koog Kim[a]

*National Creative Research Initiative Center for Spin Dynamics and Spin-Wave Devices, Nanospinics Laboratory, Research Institute of Advanced Materials, Department of Materials Science and Engineering, Seoul National University, Seoul 151-744, Republic of Korea*

We studied interactions between propagating spin waves (SWs) and a single skyrmion in a perpendicularly magnetized CoFeB nanostripe where the magnetic layer is interfaced with W and MgO. Micromagnetic numerical calculations revealed that robust interactions between the incident SWs and the skyrmion give rise to considerable forward skyrmion motions for specific SW frequencies (e.g., here: $f_{sw}$ = 12 – 19 GHz). Additionally, it was found that there exists a sufficiently low threshold field amplitude, e.g., 0.1 kOe for the $f_{sw}$ = 15 GHz SWs. This frequency-dependent interaction originated from the robust coupling of the SWs with the internal modes of the skyrmion, through the SWs' linear momentum transfer torque acting on the skyrmion. This work provides for all-magnetic control of skyrmion motions without electronic currents, and facilitates further understanding of the interactions between magnons and topological solitons in constricted geometries.

[a] Author to whom all correspondence should be addressed. Electronic mail: sangkoog@snu.ac.kr

Magnetic skyrmions, which are topological solitons with an integer topological charge,[1-3] have been found in magnetic bulk materials of non-centrosymmetry[3,4] as well as in magnetic thin films with large spin-orbit coupling at inversion-symmetry-broken interfaces.[5,6] It is known that this anti-symmetric coupling, known as Dzyaloshinskii-Moriya (DM) interaction,[7] plays a crucial role in stabilizing skyrmion formation.[3-6] Skyrmions' topological stability is advantageous in applications to memory devices, owing to both their nano-scale dimensions and ultra-low critical current density.[6, 8-11] In this light, reliable manipulation of magnetic skyrmions by spin-polarized currents or magnetic fields has attracted great interest.[6, 8-14] Very recently, skyrmion motions also have been found to be driven by spin waves (SWs) propagation in nanostripes.[15,16] This alternative approach is of special interest in terms of the promise of all-magnetic control of skyrmions in geometrically constricted elements. However, the underlying physics of SW-skyrmion interactions are still lacking. In the present study, we employed micromagnetic numerical simulations to examine SW-driven skyrmion motions and elucidate their underlying physics.

As our model, we utilized a perpendicularly magnetized CoFeB thin nanostripe of the following dimensions (see Fig. 1(a)): length $l$ = 1002 nm, width $w$ = 102 nm, and thickness $h$ = 1 nm. The CoFeB layer is interfaced with W and MgO in order to employ the Dzyaloshinskii-Moriya interactions (DMIs)[17,18] and perpendicular magnetic anisotropy,[17,19] respectively. The intended initial state is a single skyrmion with its core's out-of-plane magnetization pointing in the -$\hat{\mathbf{z}}$ direction. In our micromagnetic calculations, we employed the OOMMF code (version 1.2a5),[20] which utilizes the Landau-Lifshitz-Gilbert (LLG) equation[21] expressed as $d\mathbf{M}/dt = -\gamma/(1+\alpha^2)\mathbf{M}\times\mathbf{H}_{\text{eff}} + (\alpha/M_s)[\mathbf{M}\times(\mathbf{M}\times\mathbf{H}_{\text{eff}})]$, where $\alpha$ is the damping constant, $\gamma$ the gyromagnetic ratio, $\mathbf{H}_{\text{eff}}$ the effective field, and $M_s$ the saturation magnetization. The material parameters for the W/CoFeB/MgO model system were as follows:[17] $M_S = 1.0 \times 10^6$ A/m, exchange stiffness $A_{ex} = 2.0 \times 10^{-11}$ J/m, $\alpha$ = 0.015, perpendicular anisotropy constant $K_u = 0.8 \times 10^6$ J/m$^3$, and DMI constant $D = 1.8 \times 10^{-3}$ J/m$^2$. The unit cell size was $2 \times 2 \times 1$ nm$^3$, which is smaller than the exchange length ($l_{ex}$ = 5.6 nm) and DMI length ($l_D$ = 10 nm). In order to examine the SW interaction with the single skyrmion presented at the center, we excited monochromatic SWs by a sinusoidal magnetic field $\mathbf{H} = H_0 \sin(2\pi f_H t)\hat{y}$ of field frequency $f_H$ and field amplitude $H_0$, where the SW frequency $f_{sw}$ is equal to $f_H$.[22] The magnetic field was applied only to a local region on the left edge, denoted by the gray color in Fig. 1(a). In order to prevent reflection of SWs from the right end, we applied an abrupt absorbing boundary by setting $\alpha$ = 1 at the right-end cells.[23]

The plan-view serial snapshot images shown in Fig. 1(b) represent the Skyrmion displacements by incident SWs for different field frequencies, here $f_H$ = 5, 15, and 25 GHz, with a given value of $H_0$ = 1 kOe. The 15 GHz SWs allow for a relatively large skyrmion displacement in the same direction as the SW propagation, while the 5 GHz SWs lead to no motion, and the 25 GHz SWs, to a small extent of motion. Figure 2(a) plots the skyrmions' velocities and displacements (insets) versus time for $f_{sw}$ = 5, 15, and 25 GHz with $H_0$ = 1 kOe. As indicated, the velocities contrast markedly for the different $f_{sw}$ values. For the incident 15 GHz SWs, the skyrmion velocity increases with time and reaches a certain value, about 2 m/s. Although this velocity is low, it is yet sufficiently high for the 15 GHz SWs. For the case of $f_{sw}$ = 15 GHz, we conducted further simulations, varying the $H_0$ in the $H_0$ = 0.1 – 2 kOe range, in order to determine the effect of $H_0$ on the skyrmion velocity. The results, plotted in Fig. 2(b), show that the velocities increase with $H_0$ and then saturate to certain values that change with $H_0$. For the relatively large value of $H_0$ = 2 kOe for example, the velocity decreases with time after reaching a certain value (here, 4 m/s). This likely was due to the SWs' amplitude attenuation and edge repulsion upon the skyrmion reaching the nanostripe edge.[15] Also, we note there exists a corresponding threshold $H_0$ value to drive a given skyrmion motion (e.g., $H_0$ = 0.1 kOe for $f_{sw}$ = 15 GHz).

In order to elucidate the underlying physics of the observed frequency-dependent skyrmion motions, we performed additional micromagnetic simulations, this time applying a sinc-function field $\mathbf{H} = H_0 \sin(2\pi f_H t)/(2\pi f_H t)\hat{y}$ with $H_0$ = 10 Oe and $f_H$ = 100 GHz for the same nanostripe, with and without a single skyrmion at the center. From Fast Fourier Transformations (FFTs) of the temporal $m_z$ oscillations at each position $x$ in the longitudinal direction, we obtained the frequency spectra and corresponding dispersion curves shown in Figs. 3(a) and 3(b), respectively, for both cases without (left column) and with (right) a single skyrmion. From the comparison of the $f_{sw}$-$x$ spectra between the only-nanostripe and single-skyrmion cases, the existence of the internal skyrmion mode was evident in the $f_{sw}$ = 12 – 19 range, while a forbidden band below $f_{sw}$ = ~10 GHz, originating from the width confinement of the nanostripe (as discussed in Ref. 24), was clearly seen for both cases. Owing to the internal modes, strong reflections of the incident SWs of specific $f_{sw}$ = 12 – 19 GHz frequency range also occur, as shown in the comparison of both dispersion curves within the $k_x$ < 0 range. The very weak higher branches, namely the transverse SW modes, also are shown in both nanostripes, as reported from one of our earlier studies.[25,26]

To correlate the $f_{sw}$-dependent velocity behavior with the skyrmion internal modes, we distinguished the internal modes of a single skyrmion in the given nanostripe. To that end, we performed additional simulations by exciting, with $H_0$ =1 kOe, monochromatic SWs of different single frequencies within the 5 – 95 GHz range in increments of 0.5 GHz for 10 – 30 GHz and of 5.0 GHz for 30 – 100 GHz. The internal skyrmion modes were obtained from the FFTs of the temporal evolution of the local $m_z$ values averaged over the skyrmion region representing the isolation of the skyrmion from the remaining spin textures in the nanostripe.[9,10] Figure 4 compares the velocity averaged over $\Delta t = 100$ ns as a function of $f_{sw}$ with the FFT power of the internal skyrmion modes. As is apparent, the overall shapes of the curves are in good agreement. For $f_{sw} < 11$ GHz, SWs propagations are forbidden, due to the lateral width confinement.[24] The velocity in the low frequency range of $f_{sw} = 12 - 19$ GHz shows a quantitative agreement with the FFT power spectrum of the internal skyrmion modes, along with two major peaks at $f_{sw} = 12.5$ and 16 GHz. This reflects the fact that the frequency-sensitive skyrmion motion is closely related to the internal modes. An earlier report on domain-wall (DW) dynamics already indicated that SWs can lead to considerable DW motions for cases where SW frequencies coincide with the internal DW modes.[26] We anticipate that, in a similar manner, considerable SW reflection from the skyrmion will occur in the specific 12 – 19 GHz range, which corresponds to the skyrmion internal modes. By the resonant transfer of SWs' linear momentum to the skyrmion, the following two effects occur: SWs reflect from the skyrmion, and, simultaneously, the transfer torque drives forward skyrmion motion in the same direction as the SW propagation.[26-29] Contrastingly, SWs also are transmitted though the skyrmion, which results in a backward motion similar to the SW-driven DW motions reported in Refs 28, 29, and 30. This backward skyrmion motion, though, seems to result in much slower net motions than the SW-driven DW ones. Zhang *et al.*[15] have already reported, from an application point of view, that SWs can enable skyrmion motions in various constricted geometries including nanotracks, L-corners, T- and Y-junctions. However, until now, fundamental understanding has been lacking. In the present work, we definitively established why specific SW frequencies allow for higher-speed skyrmion motions than do other frequencies, by examining the internal modes of the skyrmion present in the nanostripe and their resonant interaction with propagating SWs. To achieve much faster forward skyrmion motion, it would be necessary to find a way to stimulate much stronger interaction.

In summary, we observed considerable forward skyrmion motions driven by SWs of specific frequencies. The motion velocity varies with the incident SWs' frequency and amplitude. For the given dimensions of a W/CoFeB/MgO nanostripe, the skyrmion is driven well by SWs of 12 – 19 GHz frequency, which frequency band corresponds to the single skyrmion's internal modes. In this frequency band, SWs' linear momenta are resonantly transferred to the skyrmion, thus resulting in linear momentum transfer torques on the skyrmion. This work provides clear physical insight into the robust interaction between SWs and a skyrmion, by which mechanism, efficient all-magnetic control of skyrmion motions in geometrically confined magnetic elements is achievable. This work would also serve as an indispensable guide to scientists and engineers, both experimentally and as relates to implementation of real applications.

This research was supported by the Basic Science Research Program through the National Research Foundation of Korea (NRF) funded by the Ministry of Science, ICT & Future Planning (Grant No. 2014001928).

**Figure captions**

FIG. 1. (Color online) (a) Model geometry and initial spin configuration including single skyrmion located at center of given nanostripe of indicated dimensions. The colors, as indicated by the color bar, correspond to the out-of-plane components of the local magnetizations $m_z = M_z / M_s$. Spin waves (SWs) are excited from the left edge (marked by the grey color). (b) Plane-view snapshot images of skyrmion motions driven by incident SWs of different pumping frequencies $f_H$ = 5, 15, and 25 GHz along with constant field amplitude $H_0$ = 1.0 kOe

FIG. 2. (Color online) Skyrmion velocity vs. time for (a) different $f_H$ values 5, 15, 25 GHz with $H_0$ = 1.0 kOe and (b) different field amplitudes $H_0$ = 0.1, 0.25, 0.5, 1.0, and 2.0 kOe with $f_H$ = 15 GHz. Each inset shows the corresponding skyrmion displacements vs. time. From the corresponding displacements, the skyrmion velocities for 100 ns were obtained.

FIG. 3. (Color online) (a) Frequency spectra along $x$ axis, as obtained from FFTs of $m_z$ oscillations averaged over nanostripe width for cases without (left column) and with (right column) skyrmion. (b) Corresponding dispersion ($f_{sw}$-$k$) curves.

FIG. 4. Comparison of FFT power vs. frequency (green solid line, representing internal modes inherent in skyrmion) with average velocity of skyrmion motion vs. frequency (blue open circle). The inset represents the region in which the $m_z$ fluctuation was spatially averaged to obtain the internal modes.

**Figures**

**Fig. 1**

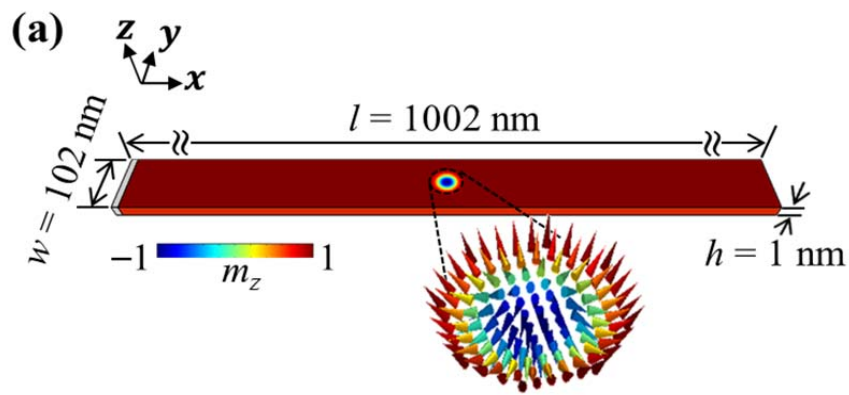

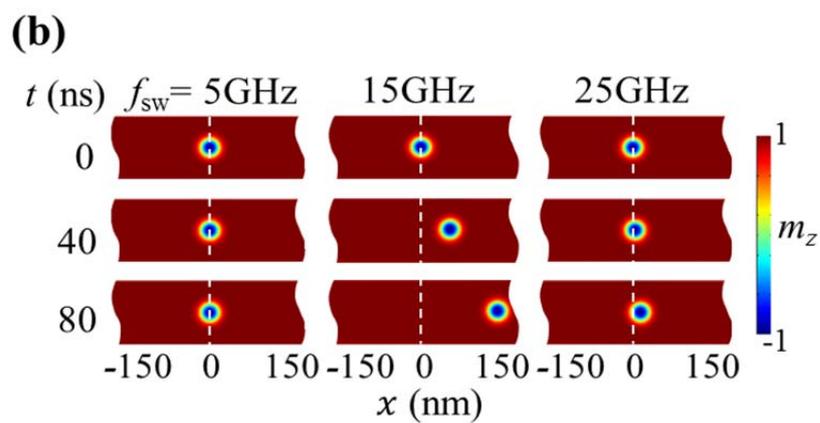

**Fig. 2**

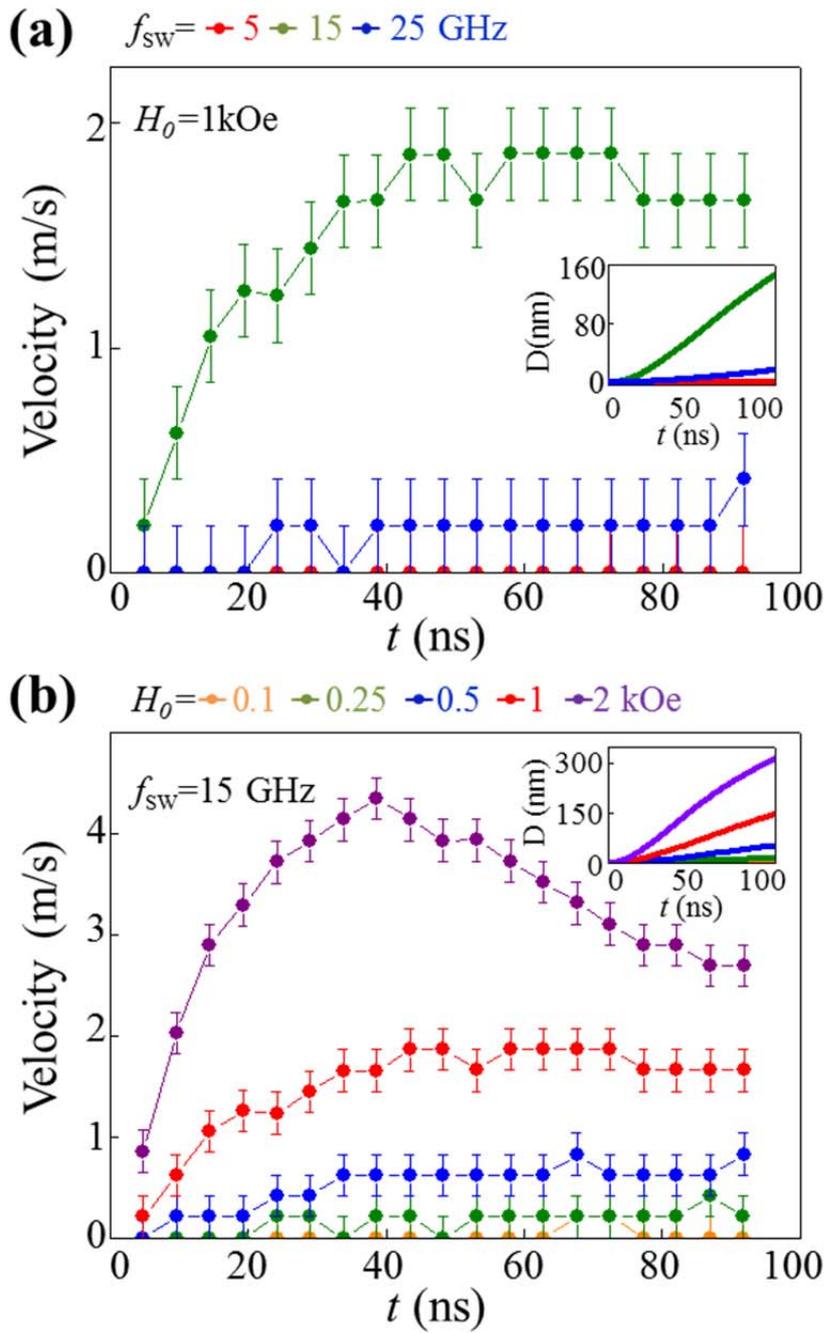

**Fig. 3**

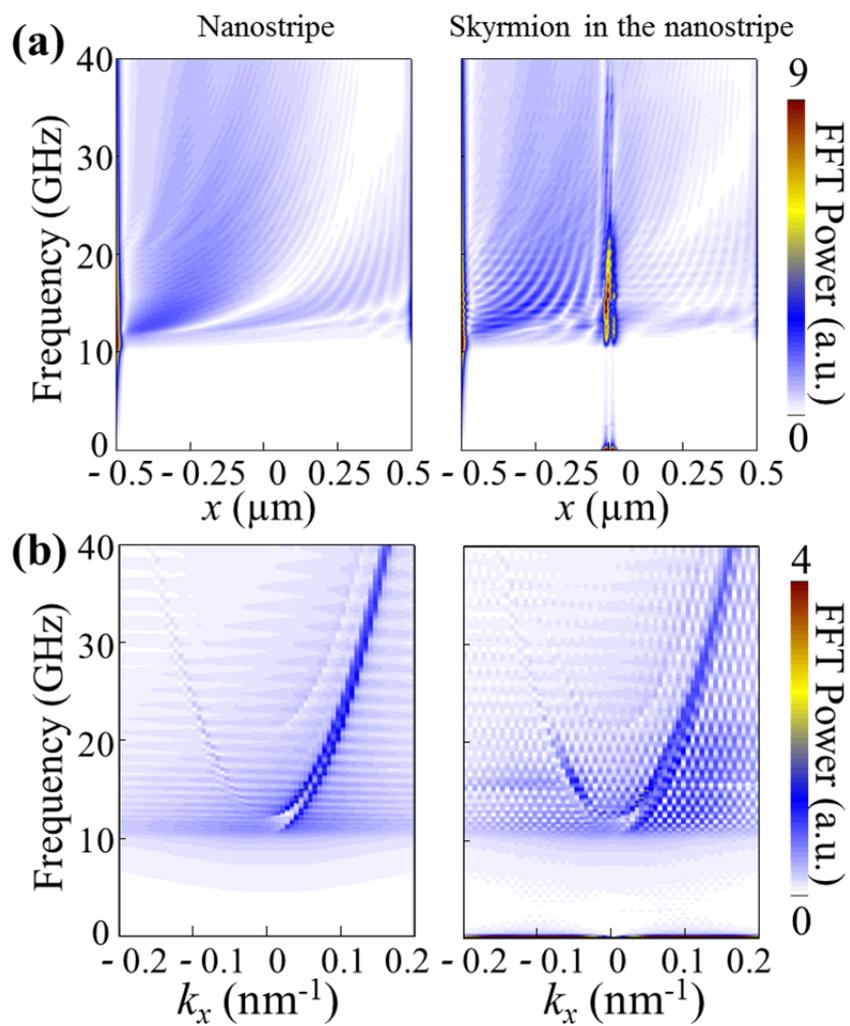

**Fig. 4**

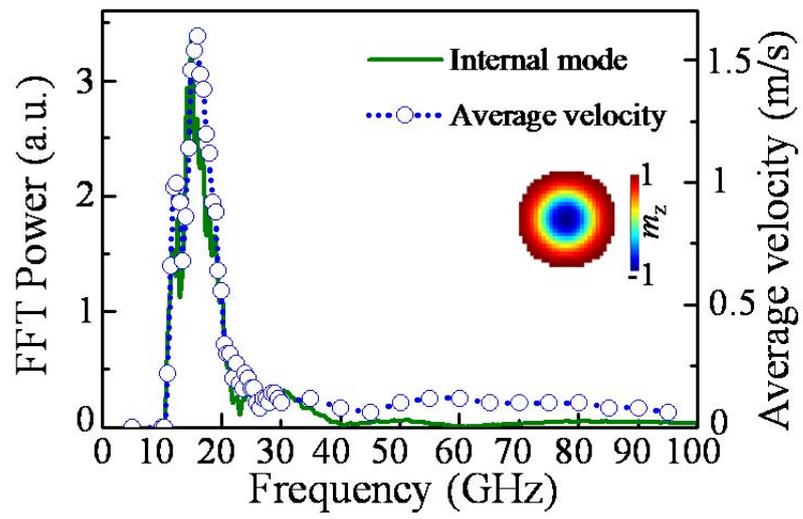